\documentstyle[11pt,oldlfont]{article}
\def\Re{\rlap{\rm I}\mkern3mu{\rm R}}
\def\a{\alpha}

\def\d{\delta}

\def\f{\varphi}

\def\k{\kappa}
\def\l{\lambda}

\def\m{\mu}

\def\r{\rho}

\def\Hb{{\cal H}}

\def\Bb{{\cal B}}
\def\Mc{{\cal M}}

\def\st{\mbox{\boldmath$*$}}

\def\S{\mbox{\boldmath$\Sigma$}}

\def\X{\mbox{\boldmath$\Theta$}}
\def\M{\mbox{\boldmath$\Xi$}}
\def\Mb{\mbox{\boldmath$\not \Xi$}}
\def\Ab{\mbox{\boldmath$A$}}
\def\Kb{\mbox{\boldmath$K$}}
\def\Yb{\mbox{\boldmath$Y$}}
\def\xib{\mbox{\boldmath$\xi$}}
\def\kb{\mbox{\boldmath$\k$}}

\def\Eb{\mbox{\boldmath$E$}}
\def\fb{\mbox{\boldmath$\f$}}

\def\p{\partial}

\def\th{\mbox{\boldmath$\theta$}}

\def\w{\mbox{\boldmath$\omega$}}
\def\we{\mbox{\footnotesize \boldmath$\wedge$}}
\def\L{\mbox{\boldmath$L$}}

\def\J{\mbox{\boldmath$J$}}
\def\U{\mbox{\boldmath$U$}}

\def\Wb{\mbox{\boldmath$W$}}

\def\Db{\mbox{\boldmath$\Delta$}}

\def\i{\mbox{\boldmath$i$}}
\def\R{\mbox{\boldmath$R$}}
\def\li{\mbox{$\mathcal{L}$}}
\def\ex{\mbox{\boldmath$d$}}
\def\cex{\mbox{\boldmath$D$}}

\def\bi{\bibitem}

\begin{document}

\thispagestyle{empty}

\begin{flushright}  
                    LPTENS 00/22  \\
                     AEI-2000-032  \end{flushright}

\vspace*{0.2cm}

\begin{center}{\LARGE {Currents and Superpotentials in classical
gauge theories: II. Global aspects and the example of Affine gravity}} 

\vskip0.2cm

B. Julia$^a$  and S. Silva$^b$

\vskip0.2cm

$^a$Laboratoire de Physique Th{\'e}orique CNRS-ENS\\ 
24 rue Lhomond, F-75231 Paris Cedex 05, France\footnote{UMR 8549 du
CNRS et de l'{\'E}cole Normale Sup{\'e}rieure.
This work has been partly supported by the EU TMR contract
ERBFMRXCT96-0012.
}\\
\vskip0.2cm
$^b$Max Planck Institut f{\"u}r Gravitationsphysik, Albert Einstein Institut,\\
Am M{\"u}hlenberg 5, D-14476 Golm, Germany

\vskip0.2cm

\begin{minipage}{12cm}\footnotesize

{\bf ABSTRACT}
\bigskip

The conserved charges associated to gauge symmetries are 
defined  at a  boundary  component of space-time because
the corresponding Noether current can
be rewritten on-shell as the divergence of a superpotential. 
However, the latter is afflicted by ambiguities. 
Regge and Teitelboim found a procedure to lift the arbitrariness
 in the Hamiltonian framework.
An alternative covariant formula  was proposed by one of us 
for an arbitrary variation of the 
superpotential, it depends only on the equations of motion and on the 
gauge symmetry under  consideration.
Here we emphasize 
that in order to compute the charges, it is enough to stay at a boundary 
of 
spacetime, without requiring any hypothesis about the bulk or about
other boundary components, so one may speak of holographic charges.
It is well known that the asymptotic symmetries that lead to conserved 
charges are
really defined at infinity, but the choice of boundary conditions and surface
terms in the action and in the charges is usually 
determined through integration by parts whereas 
each component of the boundary should be considered separately.  
We treat the example of gravity (for any space-time dimension, with 
or without cosmological constant), 
 formulated as an Affine theory which
is a natural generalization of the Palatini and Cartan-Weyl (vielbein) first 
order formulations. 
We then show that the superpotential associated to a  Dirichlet boundary
condition on the metric (the one needed to treat asymptotically flat or AdS 
spacetimes) is the one proposed by Katz, Bi\u{c}{\'a}k
and Lynden-Bell and not that of Komar. 
We finally discuss the KBL superpotential at null infinity.

\bigskip
\end{minipage}
\end{center}
\newpage

\section{Introduction}

Any Noether current associated to a gauge symmetry can be rewritten on-shell
as the
divergence of a superpotential. This is a very general property  and does not
 depend
on the gauge invariant theory we are considering \cite{JS1}.
Let us be more precise. Suppose that a gauge symmetry variation is given 
locally by
{\rm
\begin{equation}\label{symtran}
\d_\xi \fb^{i} = \ex \xi^\a \we \Db_\a^{i} + \xi^\a \we \tilde{\Db}_\a^{i}
\end{equation}
where $\fb^{i}$ denotes a $p_{i}$-form field. The
$(p_{i}-1)$ and $p_{i}$ space-time forms,  $\Db_\a^{i}$ and
$\tilde{\Db}_\a^{i}$, are functions of the fields.
Then the Noether current associated to  a  
one parameter symmetry subgroup (\ref{symtran})
is {\it always} given by:
\begin{equation}\label{fpst}
\J_\xi = \ex \U_\xi + \Wb_{\xi}
\end{equation}
with the definition
\begin{equation}\label{defww}
\Wb_{\xi}:=\xi^\a \Db_\a^{i} \we \Eb_{i},
\end{equation}
where the $(D-p_{i})$-forms $\Eb_{i}:=\frac{\d \L}{\d \fb^{i}}$ are the
Euler-Lagrange equations
associated to the field $\fb^{i}$. We used equation (\ref{symtran}) in the
definition (\ref{defww}). Note that
the {\it Noether equation} $\ex \J_\xi=  \d_\xi \fb^{i} \we \Eb_i$ follows 
from
(\ref{fpst}) and from the {\it Noether identities} due
to the gauge symmetry, namely, $\ex \Wb_\xi = \d_\xi \fb^{i} \we \Eb_i$.
The current is associated to a one-parameter subgroup of ``rigid''
 symmetries
when $\xi^{\alpha}$ can be globally taken as constant or at least has a canonical 
spacetime dependence as in the case of rotations, but the form (\ref{fpst}) 
follows from local gauge invariance along that one parameter subgroup.

In our previous work \cite{JS1}, we emphasized the fact that the Noether 
method does not
define the current $\J_\xi$ and its superpotential $\U_{\xi}$ unambiguously. 
This is
the well-known fact that the Noether current is defined up to some exact 
form,
namely $\J_\xi \sim\J_\xi+\ex \Yb  $. This exact term may contribute  
to the Noether charge when the space-time has some boundary with 
non-vanishing fields on it.
A {\it case by case} prescription
which depends on the boundary conditions is then needed in order to define
$\J_\xi$. An attempt to give a general formula for $\U_{\xi}$
can give
rise to incorrect results; see for instance our remarks in the case of 
supergravities \cite{HJS}.
The same problem arises in the covariant symplectic phase space
formalism. In that case, the ambiguity on the symplectic form can be fixed by
some ``covariant'' criterion as in \cite{Wa}.

In \cite{Si}, a way to solve
the ambiguity in the Noether current was proposed. The main result was
to give a formula for an {\it arbitrary} variation of the
superpotential, namely:

\begin{equation}\label{defu}
\int_{\footnotesize \Bb_{r}}\d \U_\xi = - \int_{\footnotesize \Bb_{r}} \delta \fb^{i} \we \frac{\partial \Wb_\xi}{\partial \ex \fb^{i}}
\end{equation}

We assume that our spacetime $\Mc$ is bounded by a set of $n$
$(D-1)$-dimensional time-like (or null) hypersurfaces denoted by
$\Hb_{r}$, $r=\{1,\dots ,n \}$. Then, $r$ labels the connected
time-like (or null) components of $\partial \Mc$. We denote by
$\Sigma_{t}$ a space-like Cauchy hypersurface at fixed time $t$. Then,
the closed $(D-2)$-dimensional manifold  $\Bb_{r}$ of (\ref{defu}) is
defined by 
$\Bb_{r}=\Sigma_{t}\cap \Hb_{r}$, for some $r=\{1,\dots ,n \}$ (and
then $\partial \Sigma_{t} = \sum_{r=1}^{n} \Bb_{r}$).
In our gravitational example, we will choose
$\Bb_{r}$ to be 
spatial infinity at fixed time $\Bb_{\infty}$.

It is important to recall that the formula (\ref{defu}) holds 
only if the theory has been rewritten in a {\it first order
formalism}\footnote{ About the possibility to construct a first order
theory from a second order one preserving the symmetries, see
\cite{JS2}.}, in the sense
 that both, the equations of motion and the symmetries (\ref{symtran}), 
depend at most on the first derivatives of the fields \cite{Si}. 
Note that this definition {\it does not imply} that a $1^{st}$ order
Lagrangian is {\it linear} in the first derivatives of the fields. The
simplest example of a $1^{st}$ order theory {\it quadratic} in the first
derivatives of the fields is the $5$-dimensional (abelian)
Chern-Simons Lagrangian $\L_{CS} = \Ab \we \ex \Ab \we \ex \Ab$.

The formula for the superpotential (\ref{defu}) is non-ambiguous
because it depends only on the equations of motion and on the functional
form of the gauge symmetry (\ref{symtran}) of the theory. It expresses an
arbitrary variation of the superpotential to be integrated on 
$ \Bb_{r}$ (for any $r=\{1,\dots ,n \}$) taking into account the chosen boundary conditions. No
additional information on the behavior of the fields outside of $\Hb_{r}$
is needed for {\it consistency}. In fact, equation (\ref{defu}) can be
justified using only 
the value of the fields (and their derivatives) at $\Hb_{r}$, no matter what happens in the
bulk or on the
other boundaries $\Hb_{s}$, $s\neq r$. This is indeed expected from
our experience: The ADM mass gives the total mass at spatial infinity,
independently of the number of black holes inside the spacetime (and
then independently of the boundary conditions used to describe their
horizons). This has to be contrasted with
the Hamiltonian Regge and Teitelboim procedure \cite{RT}. There, the
requirement of {\it differentiability} of the generators of first class
constraints implies that the Hamiltonian version of equation
(\ref{defu}) has to be satisfied on {\it every} $\Bb_{r}$. The same
stronger condition is also needed in the so-called covariant phase space
symplectic formalism for consistency (see for example
\cite{WCZ,ABR,Wa}).

The purpose of this paper is to use formula (\ref{defu}) to 
compute the superpotentials associated to general
relativity\footnote{The examples of Yang-Mills, higher dimensional
Chern-Simons and supergravity theories were respectively studied in
\cite{Si} and \cite{HJS} (see also \cite{SiT} for a review).}, with or without 
cosmological constant, in {\it any} spacetime dimension ($D \geq 3$).
 As stated already in \cite{JS1} (using there a case by case prescription) we 
find that the superpotential associated to a ``Dirichlet''
 boundary condition on 
the metric is the one proposed by Katz, Bi\u{c}{\'a}k and Lynden-Bell 
\cite{KBL}. The superpotential associated to Dirichlet boundary conditions on 
the connection (ie Neumann condition 
on the metric), is {\it one half} (in four spacetime dimensions) the famous 
Komar \cite{Ko} superpotential\footnote{This result was rediscovered in \cite{CN}.}. 
We use here the $gl(D,\Re)$ first order formalism developed in \cite{JS1}.
The motivation is that the {\it Affine}-$gl(D,\Re)$ formalism 
generalizes both the Palatini (dear to relativists) and the tetrad-vielbein 
(needed for supergravities) formalisms. So a computation at the 
$gl(D,\Re)$-level can be pulled back to either of these two well-known 
formalisms without much additional work.

We would like to insist on the simplicity and on the absence of any ambiguity 
or additional criteria of our present derivation of these 
gravitational superpotentials. In particular, the (general)
covariance of our results is automatic and does not have to be required
by hand.

We finally comment on the use of the KBL superpotential at null infinity.

For another approaches to compute superpotentials that do not emphasize
the boundary conditions see \cite{BCJ,AT,BFF,GMS}.

\section{The $gl(D,\Re)$ formalism for gravity and the associated superpotentials}\label{gldr}

\subsection{The $gl(D,\Re)$ gravity}\label{glgrav}

The basic idea of the $gl(D,\Re)$ formalism is to unify the two known first 
order formulations of General Relativity, namely the Palatini and the vielbein
 (orthonormal frames) formulations, both will follow from partial extremisations of our action. Let us recall the results of \cite{JS1}.

In the Palatini case, the {\it torsionless condition} for the
connection (namely $\Gamma^{\mu }_{\nu \rho }=\Gamma^{\mu }_{\rho \nu }$) is
{\it assumed} from the beginning. The {\it metric compatibility} of
the connection with the metric ($\nabla_{\mu} g_{\nu \rho} =0$)
and the Einstein equations are derived from the variational
principle. 

On the other hand, the vielbein formulation {\it assumes} the
{\it metric compatibility condition} between the flat Minkowski
metric and the associated orthonormal connection, that is, $D_{\mu}
\eta^{ab}=\partial_{\mu} \eta^{ab} + \omega_{\mu c}^{a}\eta^{cb} + \omega_{\mu c}^{b}\eta^{ac}=2\omega_{\mu}^{(ab)}=0$.
The {\it
torsionless condition} ($D_{[\mu} \theta_{\nu]}^{a}=0$) and Einstein
equations follow from the 
equations of motion of the basic fields. 

The $gl(D,\Re)$ first order formulation combines both ideas in a
nice way: nothing is assumed from the beginning and the {\it metric
compatibility} and {\it torsionless condition} are derived from the equations
of motion of the connection of the linear frame bundle $gl(D,\Re)$ (after fixing
the Projective symmetry in the Einstein gauge). The
Einstein equations are recovered as  usual.

\subsection*{\underline{The Lagrangian and the equations of motion}}

The Lagrangian of the $gl(D,\Re)$ gravity is a D-form $\L$ (D $\geq 3$ is 
 the spacetime dimension), it is a 
function of a linear 1-form connection $\w^a_{\ b}$ 
(for a Yang-Mills type $gl(D,\Re)$), of the canonical 1-form $\th^{a}$ 
($\Re^D$ valued) and of the metric $g^{ab}$ (which will be used to lift and 
lower the $\Re^D$-valued indices), as well as their first derivatives:

\begin{equation}
\L=-\frac{1}{4\kappa^{2}} \R^a_{\ c}  \we
\sqrt{\left|g\right|} g^{cb} \S_{ab}  \label{actgrav} 
\end{equation}

\noindent Where\footnote{In our previous paper \cite{JS1}, we used
a different notation, namely $2k=4\kappa ^{2}=16 \pi G$.} $4\kappa^{2}=16 \pi G$ and 

\begin{equation}
\S_{a_1...a_r} := \frac{1}{(D-r)!}
\epsilon_{a_1...a_rc_{r+1}...c_D} \th^{c_{r+1}}  \we...  \we \th^{c_D}
\end{equation}

\noindent $\epsilon_{a_1...a_D}$ being the Levi-Civita symbol
 ($\epsilon_{0...(D-1)}=1$).

Each field of the theory has a curvature,

\begin{eqnarray}
\R^a_{\ b} &:=& \ex \w^a_{\ b} + \w^a_{\ c} \we \w^c_{\ b}\\
\X^a &:=& \cex \th^a = \ex\th^a+\w^a_{\ b} \we \th^b \label{tors}\\
\M^{ab} &:=& \cex g^{ab} = \ex g^{ab} + \w^{ab} +\w^{ba}
\end{eqnarray}
called respectively the curvature 2-form, the torsion and the nonmetricity.

The Euler-Lagrange equations corresponding to (\ref{actgrav}) are given by:
\begin{eqnarray}
\frac{\d \L}{\d g^{ab}} &=& -\frac{\sqrt{\left| g\right|}}{8\kappa
^{2}} \left( \R^c_{\ a} \we \S_{cb} + \R^c_{\ b} \we \S_{ca}- g_{ab}
\R^{cd} \we \S_{cd}\right)  \label{eqmo1} \\ 
\frac{\d \L}{\d \th^a}  &=& -\frac{\sqrt{\left| g \right|}}{4\kappa ^{2}}  \R^{bc} \we \S_{bca}  \label{eqmo2} \\
\frac{\d \L}{\d \w^a_{\ b}} &=&-\frac{1}{4\kappa^{2}} \cex
\left(\sqrt{\left| g\right|} g^{bc} \S_{ac} \right)= -\frac{\sqrt{\left| g
\right|}}{4\kappa ^{2}}\left(  \Mb^{bc} \we \S_{ac} + \X^c \we
\S_{aec} g^{eb}  \right)  \label{eqmo3}
\end{eqnarray}
with $\Mb^{ab} := \M^{ab}-g^{ab} \frac{\M}{2}$ and $\M := \M^{ab} g_{ab}$.

\subsection*{\underline{The gauge symmetries}}

The Lagrangian (\ref{actgrav}) is invariant under three gauge symmetries:

\begin{enumerate}
\item [] 1) $\mathit{The\ local\ ``frame\ choice"\ freedom}$, parametrized by an arbitrary infinitesimal local matrix of $gl(D,\Re)$ namely, $\l^a_{\ b} = \l^a_{\ b} (x)$. The variations of the fields are: 

\begin{eqnarray}
\d_\l \th^a &=& \l^a_{\ b} \th^b \nonumber\\
\d_\l g^{ab} &=& \l^a_{\ c} g^{cb} + \l^b_{\ c} g^{ac} \nonumber\\
\d_\l \w^a_{\ b} &=& -\cex \l^a_{\ b}=- \ex \l^a_{\ b} - \w^a_{\ c} \l^c_{\ b}
 +  \w^c_{\ b} \l^a_{\ c}\label{glsy} 
\end{eqnarray}

\item [] 2) $\mathit{The\ diffeomorphism\ invariance}$, parametrized by an arbitrary infinitesimal vector field $\xi^\r =\xi^\r (x)$:

\begin{equation}\label{lie}
\d_\xi \th^a = \li_\xi \th^a
\end{equation}

and so on for $g^{ab}$ and $\w^a_{\ b}$. Here $\li_\xi$ is the usual Lie derivative acting on forms and is given by $\li_{\xi} =\ex \cdot \i_{\xi}  +  \i_{\xi} \cdot \ex$.

\item [] 3) $\mathit{The\ Projective\ Symmetry}$, parametrized by an arbitrary infinitesimal one-form $\kb=\kb (x)$:

\begin{eqnarray}\label{proj}
\d_\k \th^a &=& \d_\k g^{ab} = 0 \nonumber\\
\d_\k \w^a_{\ b} &=& \kb\ \d^a_{\ b}
\end{eqnarray}

\end{enumerate}

\subsection*{\underline{Palatini and vielbein formalisms}}

As was shown in  \cite{JS1}, the new feature here is that we need to fix the
Projective symmetry (\ref{proj}) in what we called the Einstein gauge
to recover 
the torsionless and metricity conditions from the equations of motion
of the linear connection (\ref{eqmo3}). The physics, namely Einstein
equations, does not depend on this gauge choice.

Now, the Palatini formalism can be recovered after fixing {\it all} the
$gl(D,\Re)$ symmetry by the canonical choice $\theta^{\ a}_\m =\d^{\
a}_\m$. On the other hand, the Cartan-Weyl (vielbein
or orthonormal frames) formalism comes after choosing $\theta^{\ a}_\m
=e^{\ a}_\m$, $e^{\ a}_\m$ being an orthonormal frame
(i.e. $g^{ab}=\eta^{ab}$, with $\eta^{ab}$  the ordinary flat
Minkowski metric). This last choice breaks the local $gl(D,\Re)$ down to 
local\footnote{In the Euclidian case, we fix $g^{ab}=\delta^{ab}$ and
the gauge group is broken to local $so(D,\Re)$.} $so(D-1,1,\Re)$.

The above gauge fixings can be generally rewritten as $\theta^{\ a}_\m
=\bar{\theta}^{\ a}_{\mu}$, with $\bar{\theta}^{\ a}_{\mu} (x)$ being
an arbitrary given frame. The {\it residual} gauge symmetry which preserves 
such a choice is a
linear combination of a diffeomorphism and a $gl(D,\Re)$ rotation such that:

\begin{equation}\label{pre}
\li_{\xi} \bar{\th}^{\ a} + \delta_{\lambda} \bar{\th}^{\ a} = 0
\end{equation}

The above equation gives the parameter $\lambda^{a}_{\ b}$
in terms of the diffeomorphism parameter $\xi^{\rho}$ . Using now the identity $\li_{\xi}
\bar{\th}^{\ a}= \cex\xi^{a}+\i_{\xi}\bar{\X}^{a}-\i_{\xi}\w^{a}_{\ b}
\bar{\th}^{\ b}$ (where $\xi^{a}:=\xi^{\rho} \theta^{\ a}_\r$ and
$\bar{\X}^{a}$ is the torsion (\ref{tors}) with $\th^{\ a}= \bar{\th}^{\ a}$
) and
the torsionless condition, equation (\ref{pre}) becomes:

\begin{equation}\label{autequ}
\left(\i_{\xi}\w^{a}_{\ b} -\lambda^{a}_{\ b} \right) \bar{\th}^{\ b}=\cex \xi^{a}
\end{equation}

In conclusion, after fixing the $gl(D,\Re)$ gauge symmetry, the remaining
symmetry is  diffeomorphisms parametrized by $\xi^{\rho}$ 
with simultaneous $gl(D,\Re)$-rotations parametrized by $\lambda^{a}_{\ b}
[\xi]$ which satisfy (\ref{autequ}) (or equivalently (\ref{pre})).

Let us clarify the above point by a simple example:
in the Palatini case ($\theta^{\ a}_\m =\bar{\theta}^{\ a}_\m =\d^{\
a}_\m$), equation (\ref{pre}) (or  (\ref{autequ})) simply gives $\l^a_{\ b}=-\p_b \xi^a$ (see
also \cite{JS1}). Using equations (\ref{glsy}-\ref{lie}), we then
check that the {\it residual} symmetry is the usual
diffeomorphism symmetry:

\begin{eqnarray*}
\delta_{\xi,\lambda }g^{ab} &=& \li_{\xi} g^{ab}
+\lambda^{a}_{\ c} g^{cb} + \lambda^{b}_{\ c} g^{ac}\\
&=&\xi^{c} \partial_{c} g^{ab} -\partial_{c}\xi^{a}
g^{cb}-\partial_{c}\xi^{b}  g^{ac}\\ 
\delta_{\xi,\lambda} \w^{a}_{\ b} &=&  \li_{\xi}
\w^{a}_{\ b}  -\ex \lambda ^{a}_{\ b}- \w^{a}_{\ d}  \lambda ^{d}_{\ b}+\w^{d}_{\ b}\lambda ^{a}_{\ d}\\
&=& (\xi ^{d} \partial_{d} \Gamma^{a}_{cb} + \partial_{c} \xi ^{d}
\Gamma^{a}_{db} +\partial_{c}\partial_{b} \xi^{a} +
\Gamma^{a}_{cd}\partial_{b} \xi ^{d} -\Gamma^{d}_{cb} \partial_{d} \xi
^{a}) \ex x^{c}
\end{eqnarray*}
with $\w^{a}_{\ b}:=\Gamma^{a}_{cb} \ex x^{c}$.

\subsection*{\underline{The Dirichlet and Neumann boundary
conditions}}

The
stationarity of the action constructed from the Hilbert Lagrangian
(\ref{actgrav}) implies the following condition at spatial
infinity\footnote{For consistency, the variational principle is
required to be 
satisfied at least near the boundary component 
where the superpotential and the conserved
charges are computed \cite{Si}.}:

\begin{equation}\label{boundhil}
-\int_{\Sigma_{\infty}}\delta \w^{a}_{\ b}\we {\S}_{a}^{\ b} = 0
\end{equation}
where we used the compact notation:

\begin{equation}\label{tilsig}
\S_a^{\ b}:=\frac{\sqrt{\left|
g \right|}}{4\kappa ^{2}} g^{bc} \S_{ac}
\end{equation}

It is  possible to add an appropriate total derivative $\ex \Kb$ to the 
Lagrangian (\ref{actgrav}) in
order to implement a given boundary condition and solve (\ref{boundhil}). 
The two generic cases are the following:

\begin{eqnarray}
D:\ &-\int_{\Sigma_{\infty}}\delta \w^{a}_{\ b}\we \Delta \S_{a}^{\
b}  = 0&\label{diri}\\
N:\ &\int_{\Sigma_{\infty}}  \Delta  \w^a_{\ b} \we
\d \S_a^{\ b} = 0 &\label{new} 
\end{eqnarray}
and the definitions,

\begin{equation}\label{notd}
 \Delta \S_{a}^{\ b} := \S_{a}^{\ b} - \bar{\S}_{a}^{\ b}\  , \  \  \Delta  \w^a_{\ b}:=\w^a_{\ b}- \bar{\w}^a_{\ b}
\end{equation}
with $\bar{\S}_{a}^{\ b}$ and $\bar{\w}^a_{\ b}$ our chosen asymptotic
background fields\footnote{The quantity $\bar{\S}_{a}^{\ b}$ is
constructed with some chosen asymptotic background metric. For example, in the 
Palatini formalism, we use
$\theta_{\mu}^{a}=\delta_{\mu}^{a}$ and $g^{ab}=\bar{g}^{ab}$ in
definition 
(\ref{tilsig}). In the vielbein formalism, we fix at infinity the
orthonormal frame by $\theta_{\mu}^{a}= \bar{e}_{\mu}^{a}$ in the
gauge $g^{ab}=\eta^{ab}$ (see above or \cite{JS1}).}
 whose arbitrary variation vanishes, $\delta
\bar{\S}_{a}^{\ b}=\delta \bar{\w}^a_{\ b}=0$. Then, the equations
(\ref{diri}-\ref{new}) are obtained by adding respectively to the
Lagrangian (\ref{actgrav}) the following surface terms $-\ex
(\w^{a}_{\ b}\we \bar{\S}_{a}^{\ b})$ and $\ex (\Delta \w^{a}_{\ b}\we
\S_{a}^{\ b})$.

We then recognize from (\ref{diri}-\ref{new}) typical Dirichlet and
Neumann boundary conditions: we respectively fix the asymptotic
value of the metric (through $\bar{\S}_{a}^{\ b}$ ) or of the
connection (through $\bar{\w}^a_{\ b}$). Any linear combination of
(\ref{diri}-\ref{new}) is also an appropriate boundary condition,
compatible with a variational principle.

We will be mostly interested in the cases where the metric is
asymptotically flat (or AdS) at spatial infinity. In that case, the
appropriate boundary condition is given by (\ref{diri}), with 
$\bar{\S}_{a}^{\ b}$ constructed from the flat (or AdS) metric. A covariant
way to check the vanishing of (\ref{diri}) is to use the compactification of
 Ashtekar and Romano \cite{AR} for spatial infinity. 
For an asymptotically AdS space, we can use the usual compactification
of Penrose \cite{Pe} (see also the recent \cite{AsD}).
We will not give the complete proof of these
statements here. The case of null infinity is quite a bit more involved since
neither (\ref{diri}) nor (\ref{new}) are satisfied there \cite{WZ}. We
shall elaborate on this in section \ref{nullinf}.

\subsection{The associated superpotentials}

The superpotentials associated to the gauge symmetries
(\ref{glsy}-\ref{proj}) have been computed with the cascade equations
in \cite{JS1}. The background contribution for the superpotential was
however missing there. Let us emphasize that the background is only used at 
infinity, it is nothing but a covariant way to impose the required boundary 
conditions. Our purpose will now be to use
formula (\ref{defu}) to get the complete result.

The
total superpotential will receive contributions from the $gl(D,\Re)$
(parametrized by $\l^a_{\ b}$) and diffeomorphism (parametrized by
$\xi^\r$) gauge symmetries only\footnote{The Projective
symmetry (\ref{proj}) does not contribute to $\Wb_{\xi}$ 
(see definition (\ref{defww}) together
with (\ref{symtran})). This is a consequence of the fact that no gauge
field is associated to this symmetry. In other words, there is no field
whose $\kb$-transformation law is proportional to the derivative of
the gauge parameter. Hence the associated current is identically zero
\cite{JS1}.}. 
Using the equations
(\ref{eqmo1}-\ref{proj}) in the definition (\ref{defww})
we obtain:
\begin{equation}\label{gldefw}
 \Wb_{\xi,\l} = \i_\xi \th^a\ \frac{\d \L}{\d \th^a} + \left( \i_\xi \w^a_{\ b} - \l^a_{\ b} \right)\ \frac{\d \L}{\d \w^a_{\ b}} = -\R^a_{\ b} \we \i_\xi \S_a^{\ b} - \left( \i_\xi \w^a_{\ b} - \l^a_{\ b} \right) \cex \S_a^{\ b}
\end{equation}
where the shorthand notation (\ref{tilsig}) was used. And 
$\i_\xi$ denotes the interior product with the vector field
$\xi^{\rho}$.

Now from equation (\ref{defu}), the variation of the gravitational
superpotential at some boundary $\Bb_{r}$ (and in particular at
spatial infinity $\Bb_{\infty}$) is given by 

\begin{eqnarray}
\d \U_{\xi,\l} &=& \d \w^a_{\ b} \we  \i_\xi \S_a^{\ b} +\left( \i_\xi
\w^a_{\ b} - \l^a_{\ b} \right) \d \S_a^{\ b} \label{defukbl1} \\
&=& \delta \left( \left( \i_\xi
\w^a_{\ b} - \l^a_{\ b} \right) \S_a^{\ b}\right)-\i_\xi\left( \d
\w^a_{\ b} \we   \S_a^{\ b}
\right)\label{defukbl15} \\
&=& \delta \left( D_{b}\xi^{a} \S_a^{\ b}-\i_\xi ( 
\w^a_{\ b} \we \bar{\S}_a^{\ b})
\right)-\i_\xi\left( \d 
\w^a_{\ b} \we   \Delta  \S_a^{\ b} \right)\label{defukbl2} \\
&=& \delta \left(  D_{b}\xi^{a}  \S_a^{\ b}-\i_\xi \left(\Delta \w^a_{\
b} \we  \S_a^{\ b}\right)\right)+\i_\xi\left( \Delta 
\w^a_{\ b} \we  \d \S_a^{\ b}\right)\label{defukbl3}
\end{eqnarray}

The first equation (\ref{defukbl1}) follows from the criterion 
(\ref{defu}) applied to 
(\ref{gldefw}). The second one 
(\ref{defukbl15}) follows from the first one after some simple
algebraic manipulations assuming that the variation of the gauge
parameters (namely $\xi^\r$ and $\l^a_{\ b}$) vanishes. 
The last two equations (\ref{defukbl2}-\ref{defukbl3}) are constructed
from (\ref{defukbl15}) such that the last term reproduces one of the
two boundary conditions (\ref{diri}-\ref{new}). We also used the result 
(\ref{autequ}).

The next point is to integrate equation (\ref{defukbl2}) or
(\ref{defukbl3}) using the Dirichlet or Neumann  boundary condition
(\ref{diri}-\ref{new}). The equation (\ref{defukbl1}) will be
integrable iff 
\begin{equation}\label{iLag}
\i_{\xi} (\delta \w^{a}_{\ b}\we \delta \S_{a}^{\ b})=0
\end{equation}
 at spatial infinity. This integrability condition was derived
apparently for the first time in \cite{WZ} within the covariant symplectic
formalism (the term 
\begin{equation}\label{Lag}
\Omega := \delta \w^{a}_{\ b}\we \delta \S_{a}^{\ b}
\end{equation}
 is nothing but the so-called 
pre-symplectic two-form). The correspondence between the 
covariant phase space  
formalism and equation (\ref{defu}) will be given in \cite{JS4}.

\subsubsection*{\underline{The (1/2) Komar superpotential}}

Let us first use the Neumann boundary condition on the metric (Dirichlet on
the connection) given by (\ref{new})  and integrate (\ref{defukbl1}). 
If we assume that $\w^{a}_{\ b}$ approaches $\bar{\w}^{a}_{\ b}$ fast enough, 
the last two terms of (\ref{defukbl3}) vanish.
Then, up to
some global constant, the superpotential is given by:

\begin{equation}\label{komsup}
\U_{\xi}^{\mbox{\tiny \it Ko}}=  \cex_{b} \xi^{a} \S_a^{\
b}=-\frac{\sqrt{\left|g \right|}}{4\kappa ^{2}} \st (\ex \xib) 
\end{equation}
where $\xib$ is the one-form
associated to the vector field $\xi^{\rho}$ (we also used the definition
(\ref{tilsig})).

We found the so-called Komar superpotential. However,
the coefficient is not the usual one. If we compute the charge
given by the above superpotential for the Schwarzschild black hole in $D$
spacetime dimensions, we will find $\frac{D-2}{D-3} \times M$ instead
of $M$. The superpotential found by Komar \cite{Ko} was based on
another one proposed by M\o ller \cite{Mo}. This 
author rescaled by hand the expression (\ref{komsup}) to find the
correct Schwarzschild mass for $D=4$. 

Here, we are not allowed to change the natural normalization of
(\ref{komsup}), so this superpotential is not  appropriate
to compute the usual conserved charges at spatial infinity. This is not 
surprising because it must be derived
using the boundary conditions (\ref{new}) which are
incompatible with asymptotic flatness (or AdS).

\subsubsection*{\underline{The KBL superpotential}}
 
Let us now use some Dirichlet boundary conditions on the
metric (equation (\ref{diri})) to integrate equation (\ref{defukbl1}). In that
case, the last term of equation (\ref{defukbl2}) vanishes. Thus the
superpotential is given by:
\begin{equation}\label{ukbl}
\U_{\xi}^{\mbox{\tiny \it KBL}}= \U_{\xi}^{\mbox{\tiny \it
Ko}}-\i_\xi ( \w^a_{\ b} \we \bar{\S}_a^{\ b})-C^{t}
\end{equation}

Recall that equation (\ref{defukbl1}) gives the superpotential up
to some global constant $C^{t}$. A natural way to fix this constant
is to require the vanishing of the superpotential when evaluated in
the background metric $\bar{g}^{ab}$, that is $\U_{\xi}^{\mbox{\tiny
\it KBL}}[\bar{g}]=0$. We then find a $D$-dimensional version
of the superpotential proposed by
Katz,  Bi\u{c}{\'a}k and Lynden-Bell \cite{KBL}:

\begin{eqnarray}
\U_{\xi}^{\mbox{\tiny \it KBL}}&=& \U_{\xi}^{\mbox{\tiny \it
Ko}}-\U_{\xi}^{\mbox{\tiny \it
Ko}}[\bar{g}]- \i_\xi ( \Delta \w^a_{\ b} \we \bar{\S}_a^{\ b})\nonumber\\
&=&  \U_{\xi}^{\mbox{\tiny \it
Ko}}-\U_{\xi}^{\mbox{\tiny \it
Ko}}[\bar{g}]- \i_\xi ( \Delta \w^a_{\ b} \we \S_a^{\ b})\label{ufkbl}
\end{eqnarray}
where equation (\ref{diri}) (with $\delta \w^a_{\ b}= \Delta \w^a_{\
b}$) was again used in the second line.

It is straightforward to check that the Hodge-dual of the
above $(D-2)$-form is in components equal to (see \cite{JS1} for
related calculation):

\begin{equation}\label{incomp}
\ ^{\mbox{\tiny \it KBL}}U_{\xi}^{\mu \nu }= \ ^{\mbox{\tiny \it
Ko}}U_{\xi}^{\mu \nu } - \ ^{\mbox{\tiny \it Ko}}\bar{U}_{\xi}^{\mu
\nu } +S^{\mu} \xi^{\nu}-S^{\nu} \xi^{\mu}
\end{equation}
with 

\begin{equation}\label{usdef}
\ ^{\mbox{\tiny \it Ko}}U_{\xi}^{\mu \nu }=\frac{\sqrt{\left|
g\right|}}{4 \kappa ^{2}} (\nabla^{\mu} \xi^{\nu}-\nabla^{\nu}
\xi^{\mu})\  \mbox{ and } \  
S^{\mu}=\frac{\sqrt{\left|
g\right|}}{4 \kappa ^{2}} (\Delta \Gamma ^{\mu }_{\rho \sigma} g^{\rho
\sigma}-\Delta \Gamma ^{\sigma }_{\rho \sigma }g^{\mu \rho}).
\end{equation}

In general, this superpotential\footnote{The result (\ref{incomp})
without the background contribution was derived in
\cite{JS1}   (and called the Katz superpotential \cite{Ka}) using the
cascade equations techniques.}  depends on the gauge parameter
$\xi^{\rho}$, which is not arbitrary but has to be compatible with the
asymptotic boundary condition \cite{JS1}. In the case of Dirichlet
boundary conditions 
(\ref{diri}), this parameter should be an 
asymptotic Killing vector, that is $g_{\mu \nu}+\delta_{\xi}g_{\mu
\nu}\rightarrow \bar{g}_{\mu \nu}$ as $r\rightarrow \infty $.

Our derivation of the KBL superpotential is independent of
the space-time dimension, of the first order theory used (Palatini or
Cartan-Weyl), and of the presence of a cosmological
constant\footnote{If we add a cosmological constant to the Hilbert
Lagrangian (\ref{actgrav}) by $\L_{\Lambda}=\Lambda \sqrt{\left|g
\right|} \S$, the calculations
leading to the result (\ref{incomp})  remain almost unchanged.}. It
follows straightforwardly from equation (\ref{defu}) and from
Dirichlet boundary conditions on the metric. This is to be contrasted with the
more involved original derivation \cite{KBL} and our derivation in
\cite{JS1} both with the Noether method where some prescription was needed in order to fix the
ambiguity in the current $\J_{\xi}$.
 The comparison between our present
derivation and the covariant  phase space method used 
recently in \cite{CN} will be
given in \cite{JS4}. 
Moreover, the use of the Affine formalism for
gravity gives {\it directly} the term proportional to $\partial_{\mu}
\xi^{\rho}$ in the KBL superpotential which was added by hand in
\cite{CN} following some ``covariant criterion''.

Some of the properties of the KBL superpotential, and  the
validity of equation (\ref{defu}) at null infinity are discussed in
the  last section.

\section{The KBL superpotential at null infinity}\label{nullinf}

As shown in the previous section,
the KBL superpotential follows straightforwardly from the
diffeomorphism invariance of general relativity, through equation (\ref{defu}) 
and Dirichlet boundary conditions (\ref{diri}).  
Moreover, it is the {\it only} superpotential which satisfies the
following properties: 

\begin{enumerate}
\item [] $\bullet$ It is generally covariant and then can be computed
in any coordinate system.
\item [] $\bullet$ If the chosen coordinates are the 
{\it Cartesian} ones of an asymptotically flat (or AdS)
spacetime, the KBL superpotential gives the ADM mass formula \cite{ADM} (or the AD mass
\cite{AD}). 
\item [] $\bullet$ It gives the mass and angular momentum (and the
Brown and Henneaux conformal charges \cite{BH} for {\it  AdS}$_{3}$) with the
right normalization in any spacetime dimensions $D \geq 3$. More generally, it
can be used for any asymptotic Killing vector $\xi^{\rho}$.  
\end{enumerate}

We just derived the KBL superpotential at {\it spatial infinity}. The
important point is that it depends on some background metric
$\bar{g}_{\mu \nu}$. This is crucial for its general covariance.

The asymptotic background metric is not a new object of spacetime. In
fact, it fixes the boundary conditions, namely $g_{\mu
\nu} \rightarrow \bar{g}_{\mu \nu}$ as $r\rightarrow \infty$. 
It is imposed naturally at the boundary of
spacetime. It is however clear  that in general
it cannot be extended arbitrarily everywhere in
the bulk, for instance flat space may have a different topology than our 
solution sector. The choice of such a  $\bar{g}_{\mu
\nu}$ everywhere is unnatural
from the background-independent Einstein theory point of view.
Moreover, there are 
many ways to define this background metric $\bar{g}_{\mu
\nu}$ in the bulk such that its asymptotic value agrees with our
chosen boundary conditions. It is not clear why one of these choices
would be better than the others.
If we cannot properly define  $\bar{g}_{\mu
\nu}$ {\it everywhere} then we will not be able to define $\
^{\mbox{\tiny \it KBL}}U_{\xi}^{\mu \nu }$ in the bulk and interpret the
KBL superpotential as an expression for a quasi-local mass.
We believe the problem of quasi-local charges is ill posed and needs to be 
supplemented by specific boundary assumptions at the surface to be used for 
enveloping the physical object. 

Suppose now that our spacetime is also flat at future (past) null
infinity $I^{+}$ ($I^{-}$). In that case, we can uniquely extend the background
metric $\bar{g}_{\mu \nu}$ to that region. The KBL superpotential can
then be covariantly defined also on $I^{+}$. In particular, we can
integrate equation (\ref{fpst}) on a piece of $I^{+}$, namely $\Delta$ , bounded by
spatial infinity $i_{0}$ and some time-dependent cross section ${\cal
C}$ \cite{ABR}. The on-shell result is

\begin{equation}\label{bondis}
\int_{\Sigma_{\infty}} \U_{\xi}^{\mbox{\tiny \it
KBL}}=\int_{{\cal C}}  \U_{\xi}^{\mbox{\tiny \it
KBL}} + \int_{\Delta} J_{\xi}^{\mbox{\tiny \it
KBL}}
\end{equation}
with $J_{\xi}^{\mbox{\tiny \it
KBL}}=\ex \U_{\xi}^{\mbox{\tiny \it
KBL}}$ (on-shell).

It has been proved by Katz and Lerer \cite{KL} that the first term in
the rhs of (\ref{bondis}) reproduces the Bondi mass \cite{Bon}, the
Penrose linear momentum \cite{Pel} and the
Penrose-Dray-Streubel \cite{PDS} angular momentum at null infinity
(depending on the choice of the asymptotic $\xi^{\rho}$). 
The second integral in the rhs of (\ref{bondis}) is then nothing but
the total amount of charge which crossed $\Delta$. In the case where
$\xi^{\rho}$ is an asymptotic BMS translation \cite{BMS}, we recover the result
pointed out in \cite{ABR}.

As will be shown in the forecoming paper \cite{JS4}, the equation
(\ref{defu}) used to compute the superpotential is equivalent to the covariant
symplectic phase space methods \cite{WCZ,ABR,Wa}. 
In particular, it was
shown by Wald and Zoupas \cite{WZ} that equation (\ref{defu}) is non
integrable at null infinity. This is due in part to the fact that the use of equation
(\ref{defu}) is justified only if the charge is conserved
\cite{Si}, that is, if the flux of the Noether current vanishes on the
spacetime boundary $\Hb_{r}$
under  consideration. This is not the case in general for the Bondi-type charges. The point is that the equation for the variation of the superpotential can be derived from the vanishing of the flux at infinity 
provided the set of boundary conditions is ``Lagrangian'' ie provided it 
leads to the vanishing  of the symplectic form (\ref{Lag}) there.

As a final comment, it seems quite natural that a formula like
(\ref{bondis}) should also exist on the horizon $\Hb$ of a black hole. More
precisely, if we consider $\Hb$ as a boundary of spacetime, with
some boundary conditions on it, we would hope to find an associated
superpotential, and then, some conserved or Bondi-type charges depending on its
dynamical behavior. The work of Ashtekar et al. \cite{AAA}  goes
precisely in that direction. So at least for isolated  
horizons the definition of a quasi-local mass on the horizon (even in the
 nonstationary case) should go through beyond the four dimensional case they 
have considered.

\bigskip

{\bf Acknowledgments.}
We are grateful to AEI for hospitality, and to A. Ashtekar, J. Bi\u{c}{\'a}k and
 M. Henneaux for discussions.

\end{document}